\documentclass[12pt]{article}
\usepackage{amssymb,amsfonts,amsbsy}
\usepackage{hyperref}
\newcommand{\bi}[1]{\boldsymbol{#1}}

\newcommand{\be}{\begin{equation}}
\newcommand{\ee}{\end{equation}}
\newcommand{\ba}{\begin{array}}
\newcommand{\ea}{\end{array}}
\newcommand{\p}{\partial}
\newcommand{\bu}{\bi{u}}
\newcommand{\F}{\bi{F}}
\newcommand{\G}{\bi{G}}
\newcommand{\Q}{\bi{Q}}

\newcommand{\bB}{\bi{B}}
\def\ord{\mathop{\rm ord}\nolimits}

\def\ord{\mathop{\rm ord}\nolimits}

\newcommand{\ds}{\displaystyle}
\newtheorem{thm}{Theorem}

\begin{document}
\title{\protect\vspace*{-1.5cm}{\bf On the classification of conditionally
integrable evolution systems in (1 + 1) dimensions}}
\author{A. Sergyeyev\\
Silesian University in Opava, Mathematical Institute,\\ Na
Rybn\'\i{}\v{c}ku~1, 746\,01 Opava, Czech Republic\\ E-mail: {\tt
Artur.Sergyeyev@math.slu.cz}\protect\vspace{-5mm}}
\date{}
\maketitle
\begin{abstract} 
We generalize earlier results of Fokas and Liu and find all
locally analytic (1+1)-dimen\-si\-on\-al evolution equations of order $n$ that admit
an $N$-shock type solution with $N\leq n+1$.

To this end we develop a refinement of the technique from our earlier work
\cite{as-jpa}
where
we completely characterized all (1+1)-di\-men\-si\-on\-al evolution systems
$\bi{u}_t=\bi{F}(x,t,\bi{u},\p\bi{u}/\p x,\dots,\allowbreak\p^n\bi{u}/\p x^n)$
that are conditionally invariant under a given generalized
(Lie--B\"acklund) vector field $\bi{Q}(x,t,\bi{u}, \allowbreak\p\bi{u}/\p x,
\dots,\allowbreak \p^k\bi{u}/\p x^k)\p/\p\bi{u}$
under the assumption that
the system of ODEs $\bi{Q}=0$ is totally nondegenerate.
Every such conditionally invariant evolution system admits a reduction to a system of ODEs in $t$,
thus being a nonlinear counterpart to quasi-exactly solvable models in
quantum mechanics.
\looseness=-2

\medskip


\medskip


\medskip

{\bf UDC} 517.95+514.763.85
\end{abstract}

\section{Introduction}

In order to be relevant, mathematical models of natural phenomena
are often required to have solutions of  prescribed form
(e.g.\ travelling wave solutions, kinks, solitons, etc.).
As was first shown by Fokas and Liu \cite{fliu, fliu2} (see also Zhdanov
\cite{zhd}), a natural way to achieve this
for evolution systems in (1+1) dimensions is to require that the
system in question admit a generalized conditional symmetry
(GCS) such that the 
solutions invariant under this symmetry have
the desired form. It turns out that the set
of systems admitting GCS (such systems, or even more broadly,
the systems possessing integrable reductions, are often referred to as
{\em conditionally integrable}, see e.g.\ \cite{wint}) is considerably larger
than that of systems integrable via
the inverse scattering transform or direct linearization,
see e.g.\  \cite{fliu, fliu2}. Notice \cite{zhd4} that
if an (1+1)-dimensional evolution equation admits a reduction
to a system of ODEs in the evolution parameter (time $t$), then it admits a GCS,
and vice versa.
\looseness=-1

This naturally leads to the following problem: how to describe all
evolution systems that admit a given GCS? For the
case of linear GCS with time-independent coefficients
this problem, restated in the terms of the
so-called invariant modules, was completely solved in the seminal
paper \cite{imod} (see also an important earlier work \cite{svirmfti}).
For the nonlinear GCS, some partial results
were obtained in \cite{andr}, and the complete solution of the
problem was obtained in \cite{as-jpa} under certain nondegeneracy
assumptions, see below for details; see also \cite{imod}
and \cite{as-jpa} for the survey of
earlier results in the field and the discussion of
the role played by the GCS in the search of exact solutions
of nonlinear PDEs.
\looseness=-1

However, the formulas found in \cite{as-jpa} are not very
convenient for applications. In the present paper we obtain
alternative, easier to use, formulas for the locally analytic
(1+1)-dimensional evolution systems 
that are conditionally invariant under a given {\em analytic} GCS.
This is done in Section 3.
In Section~4 we illustrate the application of our results by a number of examples.
In particular, we completely characterize all
locally analytic (1+1)-dimensional evolution equations of order $n$ that admit
an $N$-shock type solution with $N\leq n+1$.
\looseness=-1

\section{Preliminaries}
Consider an
evolution system
\begin{equation} \label{eveq}
 \partial \bu / \partial t =\F(x, t,\bu, \bu_{1}, \dots, \bu _{n}), \quad n \geq 0,
\end{equation}
for an $s$-component vector function $\bu=(u^1,\dots,u^s)^{T}$,
where $\bu_{l}=\partial^{l} \bu /\partial x^{l}$,
$l=0,1,2,\dots$, $\bu_{0} \equiv \bu$, and the superscript `$T$'
denotes the matrix transposition.

A smooth 
function of $x,t,\bi{u},\bi{u}_1,\bi{u}_2,\dots$ is called
{\em local} (see \cite{mik}; cf.\ also \cite{vin} and \cite{olv_eng2}) if it depends on a
finite number of $\bi{u}_j$. The greatest integer $m$ such that
$\p f/\p\bi{u}_m\neq 0$ is called the {\em order} of a differential function
$f$, and we shall write that as $m=\ord f$. If $f$ depends only on $x$ and $t$,
then we shall assume that $\ord f=0$.
Unless otherwise explicitly stated,
all functions considered below will be assumed to be local.
\looseness=-1

A generalized vector field $\mathcal{Q}=\Q\,\p/\p \bu$,
where $\bi{Q}$ is an $s$-component local vector function
is called \cite{fliu, fliu2, zhd}
a {\it generalized conditional symmetry} (GCS) for (\ref{eveq}) if
the system $\bi{Q}=0$ is compatible with (\ref{eveq}).
\looseness=-1

Eq.(\ref{eveq}) is compatible with $\bi{Q}=0$ if and only
if (cf.\ e.g.\ \cite{fliu,zhd})
\be\label{cond}
D_{t}(\bi{Q})|_{\mathcal{M}}=0,
\ee
where
$D_{t}= \p /\p t + \sum_{i=0}^{\infty} D^{i}(\F) \p/\p
\bu_{i}$ and $D= \p /\p x +
\sum_{i=0}^{\infty} \bu_{i+1} \p/\p \bu_{i}$
are total $t$- and $x$-derivatives, and
$\mathcal{M}$ is the solution manifold of
the system $\bi{Q}=0$.

In what follows we shall treat $\bi{Q}=0$
as a system of ODEs involving an extra parameter $t$, and
assume that this system is totally nondegenerate, i.e.,
the systems $D^j(\bi{Q})=0$ are locally solvable and have maximal rank
for all $j=0,1,2,\dots$, see Ch.~2 of \cite{olv_eng2} for more details
on this. Then the condition (\ref{cond}) is equivalent  \cite{olv_eng2} to the following:
there exist $s$-component local vector functions
$\boldsymbol{\eta}_{\alpha,j}$ and an integer $p$ such that
\be\label{cond1} D_{t}(\bi{Q})=\sum\limits_{\alpha=1}^s
\sum\limits_{j=0}^p \boldsymbol{\eta}_{\alpha,j} D^j(Q^\alpha).
\ee
\looseness=-1

Finding all GCS admitted by a given system (\ref{eveq})
is an extremely difficult task, whose complexity is comparable to that of
the problem 
of finding all solutions for the system in question.
Surprisingly enough, we succeeded \cite{as-jpa} in  completely solving the inverse problem of describing
all systems (\ref{eveq}) that admit a given GCS $\mathcal{Q}=\bi{Q}\p/\p\bi{u}$,
provided the system $\bi{Q}=0$, considered as a system of ODEs,
is totally nondegenerate.
\looseness=-1

Let us briefly review the results of \cite{as-jpa}.
Consider a system of ODEs
\be\label{q1}
\boldsymbol{Q}(x,t,\boldsymbol{u},\boldsymbol{u}_1,\dots,\boldsymbol{u}_k)=0,
\ee
involving  $t$ as a parameter. Here $\boldsymbol{Q}=(Q^1,\dots,Q^s)^T$
is an $s$-component local vector function.

Assume that the general solution of (\ref{q1})
in implicit form can be written as
\be\label{int}
\G(x,t,\bu,c_{1}(t),\dots,c_{N}(t))=0. \ee
The number $N$ is sometimes called the total order of (\ref{q1}).
It is tacitly assumed here that $\G$ depends on $N$ arbitrary
functions $c_{i}(t)$, $i=1,\dots,N$, in an essential way, and that
$\det\p\G/\p\bu\neq 0$.\looseness=-1

Then using the implicit function theorem
we can, at least locally, obtain the general solution in the
explicit form as
$\boldsymbol{u}=
\boldsymbol{P}(x,t,c_1(t),\dots,\allowbreak c_N(t))$, whence
$$u_{j}^{\alpha}=\p^j
P^{\alpha}(x,t,c_1(t),\dots,c_{N}(t))/\p x^j.
$$
Using these
equations
we can express $c_i$ as functions of 
$x,t,u^1,\dots,u_{n_1-1}^1,\allowbreak\dots, u^s,\dots,u_{n_s-1}^s$
for some numbers $n_1,\dots,,n_s$:
$$c_{i}=h_{i}(x,t,u^1,\dots,u_{n_1-1}^1,\dots,u^s,
\dots,u_{n_s-1}^s),\quad i=1,\dots,N.$$

Let $\tilde{\bB}_i=-\left({\p\G}/{\p\bu}\right)^{-1}{\p\G}/{\p c_i},
\tilde{\boldsymbol{R}}=-\left({\p\G}/{\p\bu}\right)^{-1}{\p\G}/{\p
t}$, $\bB_{i}=\tilde{\bB}_{i}(x,t,\allowbreak h_{1},\dots,h_{N})$,
and $\boldsymbol{R}(x,t,\bu,\allowbreak \bu_1,\dots,\allowbreak
\bu_{r})\equiv\tilde{\boldsymbol{R}}(x,t, \allowbreak
h_1,\dots,h_N)$, i.e., $\bB_{i}$ and $\boldsymbol{R}$ are obtained
from $\tilde{\bB}_{i}$ and $\tilde{\boldsymbol{R}}$ by substituting
$h_{i}$ for $c_{i}$. Here it is under\-stood that $c_i$
are {\em not} differentiated with respect to $t$ while evaluating
$\p\boldsymbol{G}/\p t$.
\looseness=-1



Let $V$ be an open domain in the space $\mathcal{V}$ of variables
$x,t,\bu,\bu_1,\dots$, and let $W$ be the set of all points in $V$
satisfying the equations $D^j(\Q)=0$, $j=0,1,2,\dots$, considered
as algebraic equations.
\begin{thm}[\mbox{\protect{\cite{as-jpa}}}]\label{cspr}
 Suppose that
$\Q(x,t,\bu,\dots,\bu_k)=0$, considered as a system of ODEs,
is analytic on $V$, totally nondegenerate on $W$,
and has the same total order $N$ on the whole of $W$.

If $\Q=0$ is compatible with (\ref{eveq}), i.e., $\Q\p/\p\bu$ is a
generalized conditional symmetry for (\ref{eveq}),
then $\F$ on $V$ can be represented in the form
\be\label{symq2} \ba{l}
\F=\boldsymbol{R}+\sum\limits_{i=1}^{N}\zeta_{i}(t,h_{1}
,\dots,h_{N})
\bB_{i}+\sum\limits_{p=0}^{m}\sum\limits_{\alpha=1}^{s}
\boldsymbol{\chi}_{p,\alpha}(t,x,\bu,\dots,
\bu_{j_{p,\alpha}})D^{p}(Q^\alpha), \ea \ee where $m$ and
$j_{p,\alpha}$ are nonnegative integers, and $\zeta_i$ and
$\boldsymbol{\chi}_{p,\alpha}$ are smooth functions of their
arguments. \looseness=-1
\end{thm}

By construction \cite{as-jpa}, the system
$\bi{u}_t=\F$ with $\F$ given by (\ref{symq2})
admits a solution of the same form as the general solution
(\ref{int}) of $\bi{Q}=0$, that is,
$$
\bi{G}(x,t,\bi{u},c_1(t),\dots, c_N(t))=0,
$$
but 
now the $c_i(t)$ must satisfy
\be\label{redev}
dc_i/d t=\zeta_{i}(t,c_1,\dots,c_N), \quad i=1,\dots,N,
\ee
rather than be arbitrary functions of $t$.
\looseness=-1

In other words, under the assumptions of Theorem~\ref{cspr} if
$\bi{Q}\p/\p\bi{u}$ is a GCS for $\bi{u}_t=\F$, then the
substitution of the general solution (\ref{int}) of $\bi{Q}=0$ into
$\bi{u}_t=\F$ reduces the latter to the system of ODEs
(\ref{redev}). \looseness=-1

\section{Solving the inverse problem: which systems admit a given GCS?}


How to pick among the $\F$'s (\ref{symq2}) those of order $\leq n$, where $n$
is a given natural number? This can be done using the following result.
\begin{thm}\label{an}
Under the assumptions of Theorem~\ref{cspr} on $\boldsymbol{Q}$,
suppose that \be\label{ndeg} Q^\alpha=u^\alpha_{n_{\alpha}}
-g^\alpha(x,t,\tilde u), \quad\alpha=1,\dots,s,\ee
where  $g^\alpha$ are analytic functions of their arguments
and $\tilde u=(u^1,\dots,u_{n_1-1}^1, \allowbreak\dots,
u^s,\dots,u_{n_s-1}^s)$.

Then the most general locally analytic $\boldsymbol{F}$
of order $n\geq\max(\ord\boldsymbol{R}, \allowbreak\max\limits_{i=1,\dots,N}
\ord\boldsymbol{B}_i, \allowbreak\max\limits_{j=1,\dots,N}\ord h_j)$
such that (\ref{eveq}) is compatible with
$\boldsymbol{Q}=0$ can be locally written as\looseness=-1
\be\label{anaf2} \ba{l}
\hspace*{-5mm}\F=\boldsymbol{R}+\sum\limits_{i=1}^{N}\zeta_{i}(t,h_{1}
,\dots,h_{N})\bB_{i}
+\sum\limits_{\alpha=1}^s\sum\limits_{m=0}^{n-n_{\alpha}}
D^m(Q^\alpha) \boldsymbol{K}_{\alpha,m}(x,t,\tilde u, \tilde Q), \ea
\ee where $\tilde Q=(Q^1, D(Q^1),\dots,
D^{n-n_1}(Q^1),\dots,
Q^s, D(Q^s),\dots, D^{n-n_s}(Q^s))$, $N=\sum_{\alpha=1}^s n_\alpha$,
$\zeta_i$ and $\boldsymbol{K}_{\alpha, m}$ are arbitrary
locally analytic functions of their arguments,
and for $n< n_\alpha$ $\boldsymbol{K}_{\alpha, m}\equiv 0$.
\looseness=-1
\end{thm}
{\em Proof.}
We have $u^\alpha_{n_{\alpha}}=Q^\alpha+g^\alpha(x,t,\tilde u)$.
Using this equality and proceeding inductively with usage of the
formulas found at the previous steps, we obtain
$u_j^\alpha=\psi^\alpha_j(x,t,\tilde u, Q^1,\dots, D^{j-n_1}(Q^1),
\dots,\allowbreak Q^s,\dots, \allowbreak D^{j-n_s}(Q^s))$ for
$j\geq\max\limits_\alpha n_\alpha$ and similar formulas for
$\max\limits_\alpha n_\alpha>j>n_\alpha$.


Plug the above formulas for $u^\alpha_j$
into (\ref{symq2}) and expand the resulting expression in Taylor series
with respect to $D^j(Q^\alpha)$. As the order of $\boldsymbol{F}$
is $n$ and $\boldsymbol{F}$ is locally analytic, it must be independent of
$D^j(Q^\alpha)$ with $j>n-n_\alpha$.
Taking this into account, we can rewrite (\ref{symq2}) in the form
(\ref{anaf2}), and the result follows.
\looseness=-1
$\square$

It is easily seen that the most general locally analytic $\F$ of
order $n<\tilde n\equiv \max(\ord\boldsymbol{R},
\allowbreak\max\limits_{i=1,\dots,N} \ord\boldsymbol{B}_i,
\allowbreak\max\limits_{j=1,\dots,N}\ord h_j)$ such that
$\bi{u}_t=\F$ admits a given GCS $\bi{Q}\p/\p\bi{u}$ meeting the
requirements of Theorem~\ref{an} also has the form (\ref{anaf2})
with $n$ replaced by $\tilde n$ and $\bi{K}_{\alpha,m}$ and
$\zeta_i$ subjected to the extra conditions $\p\F/\p\bi{u}_j=0$ for
$j>n$.
\looseness=-2

Note that for $\bi{Q}=\boldsymbol{u}_k-\boldsymbol{g}(x,t,\boldsymbol{u},
\dots,\boldsymbol{u}_{k-1})$
we have $N=s\cdot k$, and (\ref{anaf2}) boils down to
\be\label{anaf2a} \ba{l}
\F=\boldsymbol{R}+\sum\limits_{i=1}^{N}\zeta_{i}(t,h_{1}
,\dots,h_{N})\bB_{i}\\
+\sum\limits_{\alpha=1}^s\sum\limits_{m=0}^{n-k}  D^m(
Q^\alpha) \boldsymbol{K}_{\alpha,m}(x,t,
\boldsymbol{u},\boldsymbol{u}_{1},\dots,\boldsymbol{u}_{k-1},
\Q,D(\Q),\dots, D^{n-k}(\boldsymbol{Q})).\ea \ee

\section{Examples}
{\it Example 1.}
Let
\be\label{cklin}
Q^{\alpha}=u_{n_{\alpha}}^{\alpha}-\sum_{\beta=1}^{s}\sum\limits_{j=0}^{n_{\beta}-1}
g^{\alpha}_{\beta,j}(x,t) u_{j}^{\beta},\quad \alpha=1,\dots,s, \ee
and hence $N=\sum_{\alpha=1}^{s}n_\alpha$.
Then a general solution of $\Q=0$ reads
$\bu=\sum_{i=1}^{N}c_{i}(t) \boldsymbol{f}_i(x,t)$, where
$c_i(t)$ are arbitrary functions of $t$ and
$\boldsymbol{f}_i\equiv (f_i^1,\dots,f_i^s)^T$ are linearly
independent solutions of $\Q=0$.\looseness=-1

We have \cite{sam, sam2}  \looseness=-1
$h_i=Z_i/Z$, where\looseness=-1
$$
\hspace*{-7mm} \small
 Z=\left|\begin{array}{ccccccc}f_1^1 &\dots &f_i^1 &\dots
&f_N^1\\ \p f_1^1/\p x &\dots &\p f_i^1/\p x &\dots &\p f_N^1/\p
x\\ \vdots & \vdots & \vdots & \vdots & \vdots\\ \p^{n_1-1}
f_1^1/\p x^{n_1-1} &\dots &\p^{n_1-1} f_i^1/\p x^{n_1-1} &\dots
&\p^{n_1-1} f_N^1/\p x^{n_1-1}\\ \vdots & \vdots & \vdots & \vdots
& \vdots\\ f_1^s &\dots &f_i^s &\dots &f_N^s\\ \p f_1^s/\p x
&\dots &\p f_i^s/\p x &\dots &\p f_N^s/\p x\\ \vdots & \vdots &
\vdots & \vdots & \vdots\\ \p^{n_s-1} f_1^s/\p x^{n_s-1} &\dots
&\p^{n_s-1} f_i^s/\p x^{n_s-1} &\dots &\p^{n_s-1} f_N^s/\p
x^{n_s-1}\\
\end{array}
\right|
$$
and $Z_i$ are obtained from $Z$ by replacing $\p^j
f_i^{\alpha}/\p x^j$ by $u_j^{\alpha}$. It is easily seen that
$\boldsymbol{B}_i=\boldsymbol{f}_i(x,t)$,
and $\boldsymbol{R}=\sum_{i=1}^{N} (Z_{i}/Z) \p
\boldsymbol{f}_i(x,t)/\p t$.

If $g^\alpha_{\beta,j}$ are analytic in $x$ and $t$, then by
Theorem~\ref{an}
the most general locally analytic $\bi{F}$ of order $n\geq\max\limits_i\ord Z_i/Z$
that admits a GCS $\bi{Q}\p/\p\bi{u}$ with $\bi{Q}$ (\ref{cklin}) 
locally has the form
\be\label{symlin} \ba{l}
\hspace*{-7mm}\F=\sum\limits_{i=1}^{N} {\ds\frac{Z_{i}}{Z}\frac{\p
\boldsymbol{f}_i(x,t)}{\p t}}+
\sum\limits_{i=1}^N\zeta_{i}(t,Z_{1}/Z,\dots,Z_{N}/Z)
\boldsymbol{f}_{i}(x,t)
+\sum\limits_{\alpha=1}^s\sum\limits_{m=0}^{n-n_{\alpha}}
D^m(Q^\alpha) \boldsymbol{K}_{\alpha,m}(x,t,\tilde u, \tilde Q),
\ea
\ee
where $\zeta_i$ and $\boldsymbol{K}_{\alpha, m}$ are arbitrary
locally analytic functions of their arguments, $\tilde Q=(Q^1, D(Q^1),\dots,\allowbreak
D^{n-n_1}(Q^1),\dots,\allowbreak
Q^s, D(Q^s),\dots, D^{n-n_s}(Q^s))$, $\tilde u=(u^1,\dots,u_{n_1-1}^1,\dots,
u^s,\dots,u_{n_s-1}^s)$, and $\boldsymbol{K}_{\alpha, m}\equiv 0$ for $n< n_\alpha$.
\looseness=-1


In the rest of this section we assume that $s=1$, so for simplicity we shall
write $\bu\equiv u$, $\Q\equiv Q$,~$\F\equiv
F$. 
\looseness=-1

{\it Example 2.}
Let $Q=L(u)$, where $L=\prod_{j=1}^{N}(D-k_j)$, and $k_i$, $i=1,\dots,N$,
are distinct ($k_i\neq k_j$ if $i\neq j$) nonzero constants.
Then the general solution of $Q=0$ is $\sum_{i=1}^{N}c_i(t)\exp(k_i x)$,
where $c_i(t)$ are arbitrary functions of $t$,
and using the results of Example 1 we obtain
\looseness=-1
$$h_i=
\frac{\exp(-k_i x)}{\prod\limits_{j=1,\\ j\neq i}^{N}(k_i-k_j)}
L_i(u),
$$
where $L_i=\prod_{j=1,j\neq i}^{N}(D-k_j)$.
Thus, the most general locally analytic $F$ of order $n\geq N-1$ such that $u_t=F$ admits a GCS $Q\p/\p u$ with
$Q=L(u)$ can be locally written as
\be\label{linf}
\ba{l}
\hspace*{-4.3mm}F\!=\!\sum\limits_{i=1}^N\zeta_{i}(t,h_1,\dots,h_N)
\exp(k_i x)
\!+\!\!\!\sum\limits_{m=0}^{n-N}\!\! D^m(Q) K_{m}(x,t,u,u_1,\dots,u_{N-1}, Q,
D(Q),\dots, D^{n-N}(Q)),
\hspace{-2mm} \ea \ee where $\zeta_i$ and $K_{m}$
are arbitrary locally analytic functions of their arguments, and
$K_{m}\equiv 0$ if $n<N$.

{\it Example 3.}
Let $Q=u_2-f(u,t)$, where $f$ is an arbitrary analytic function of $u$ and $t$.
Set $a(z,t)=\int f(z,t)dz$ and $\psi(y,z,t)=\int (2 a(y,t)+z)^{-1/2}dy$.
Then the general solution of $Q=0$ in implicit form reads
$\psi(u,c_1(t),t)=x+c_2(t)$, where $c_i(t)$ are arbitrary functions of $t$,
and we have $h_1=u_1^2-2 a(u,t)$, $h_2=\psi(u, z,t)|_{z=h_1}-x$.
\looseness=-1

Hence by Theorem~\ref{an} the most general locally analytic $F$ or
order $n\geq 1$ such that $u_t=F$ admits a GCS $Q\p/\p u$ with
$Q=u_2-f(u,t)$ locally has the form
\[
\ba{l}
F=-(2 a(u,t)+h_1)^{1/2}\biggl(\biggl({\ds\frac{\p\psi(u,z,t)}{\p t}}\biggr)\biggl|_{z=h_1}
-\zeta_{1}(t,h_1,h_2){\ds\int^u \frac{dy}{2(2 a(y,t)+h_1)^{3/2}}}-\zeta_{2}(t,h_1,h_2)\biggr)\\
+\sum\limits_{m=0}^{n-2}D^m(Q) K_{m}(x,t,u,u_1, Q, D(Q),\dots, D^{n-2}(Q)),
\ea
\]
where $\zeta_i$ and $K_{m}$ are arbitrary locally analytic functions
of their arguments; if $n<2$, then $K_{m}\equiv 0$ for all $m$.
\looseness=-1


{\it Example 4.}
As a somewhat more elaborated example, consider $Q=u_2-\varphi(x,t) f(u_1,t)$, where $\varphi$
and $f$ are arbitrary analytic functions
of their arguments, and $f\not\equiv 0$. Let $a(z,t)=\int dz/f(z,t)$,
$\tilde\varphi(x,t)=\int\varphi(x,t)dx$,
and let $b(y,t)$ denote a solution of the equation $a(z,t)=y$
with respect to $z$, so that $a(b(z,t),t)\equiv z$.
Then the general solution of $Q=0$ reads $u=c_1(t)+\chi(x,c_2(t),t)$, where
$\chi(x,z,t)=\int b( \tilde\varphi(x,t) +z,t)dx$, $c_i(t)$
are arbitrary functions of $t$,
and we have
$h_1=u- \chi(x,z,t)|_{z=h_2}$, $h_2=a(u_1,t)- \tilde\varphi(x,t)$.

Hence, the most general locally analytic $F$
of order $n\geq 1$ such that $u_t=F$ admits a GCS $Q\p/\p u$ with
$Q=u_2-\varphi(x,t) f(u_1,t)$  locally  has the form
\[
\ba{l}
F=\biggl({\ds\frac{\p\chi(x,z,t)}{\p t}}\biggr)\biggl|_{z=
h_2}+\zeta_{1}(t,h_1,h_2)
+\zeta_{2}(t,h_1,h_2)\biggl({\ds\frac{\p\chi(x,z,t)}
{\p z}}\biggr)\biggl|_{z= h_2}\\
+\sum\limits_{m=0}^{n-2}D^m(Q) K_{m}(x,t,u,u_1,
Q, D(Q),\dots, D^{n-2}(Q)),
\ea
\]
where $\zeta_i$ and $K_{m}$ are arbitrary locally analytic
functions of their arguments, and  if $n<2$, then $K_{m}\equiv 0$
for all $m$.
\looseness=-1

{\it Example 5.} Assume now that $Q=M(1)$, where $M=\prod_{j=1}^{N+1}(D-k_j-u)$.
The general solution of $Q=0$
has the form $-v_x/v$, where $v\equiv\sum_{i=1}^{N+1}b_i(t)\exp(k_i x)$, and
$b_i(t)$ are arbitrary functions of $t$, cf.\ \cite{fliu}.
\looseness=-2

Notice that
$-v_x/v$ actually involves only $N$ independent
arbitrary functions of $t$.
Indeed, renumbering $b_i$ if necessary, we can assume without loss
of generality that $b_{N+1}(t)\neq 0$. Then we can
rewrite $u=-v_x/v$ in the form
\be\label{nshock}
u=-D\biggl(\ln\biggr(\exp(k_{N+1} x)
+\sum\limits_{i=1}^{N}c_i(t)\exp(k_i x)\biggr)\biggr),
\ee
where $c_{i}(t)=b_i(t)/b_{N+1}(t)$, whence
\be\label{hshock}
\ba{l}
h_i=
{\ds\frac{\exp((k_{N+1}-k_i)x)M_{i}(1)\prod\limits_{j=1}^{N}
(k_{N+1}-k_j)}{M_{N+1}(1)
\prod\limits_{j=1, j\neq i}^{N+1}(k_i-k_j)}}, \quad i=1,\dots,N.
\ea
\ee
Here $M_i=\prod_{j=1,j\neq i}^{N+1}(D-k_j-u)$.

The most general locally analytic $F$ of order $n\geq N-1$
such that $u_t=F$ admits a GCS $Q\p/\p u$
with $Q=M(1)$ (and hence an $N$-shock type
solution (\ref{nshock})) locally has the form
\be\label{shockf}
\ba{l}
F=
-\sum\limits_{i=1}^N\zeta_{i}(t,h_1,\dots,h_N)
\exp(k_i x){\ds\frac{\biggl(\sum\limits_{j=1}^{N+1}
(k_i-k_j)h_j\exp(k_j x)\biggr)}
{\biggl(\sum\limits_{q=1}^{N+1}h_q\exp(k_q x)\biggr)^2}}\\
+\sum\limits_{m=0}^{n-N} D^m(Q) K_{m}(x,t,u,u_1,\dots,u_{N-1}, Q,
D(Q),\dots, D^{n-N}(Q)), \ea \ee where $\zeta_i$ and $K_{m}$ are
arbitrary locally analytic functions of their arguments, and for
$n<N$ we have $K_{m}\equiv 0$ for all $m$. This time the $h_i$ are
of the form (\ref{hshock}), and for the sake of brevity we set
$h_{N+1}=1$. The corresponding $N$-shock type solution of the
evolution equation $u_t=F$ is of the form (\ref{nshock}) with
$c_i(t)$ that satisfy (\ref{redev}).
\looseness=-1

Notice that one can readily pick among the $F$'s (\ref{shockf})
those independent of $x$ and $t$.
They are of the form
\be\label{nshockxt}
\ba{l}
\hspace*{-7mm}F=
-\sum\limits_{i=1}^N\eta_{i}
\biggl({\ds\frac{\tilde h_1^{k_{N+1}-k_{N}}}
{\tilde h_N^{k_{N+1}-k_{1}}}},\dots,
{\ds\frac{\tilde h_{N-1}^{k_{N+1}-k_{N}}}
{\tilde h_N^{k_{N+1}-k_{N-1}}}}\biggr)
{\ds\frac{\tilde h_i\sum\limits_{j=1}^{N+1}(k_i-k_j)\tilde h_j}
{\biggl(\sum\limits_{q=1}^{N+1}\tilde h_q\biggr)^2}}\\
+\sum\limits_{m=0}^{n-N}
D^m(Q)  K_{m}(u,u_1,\dots,u_{N-1}, Q, D(Q),\dots,
D^{n-N}(Q)), \ea
\ee
where $\tilde h_i=h_i\exp((k_i-k_{N+1})x)$, and
$\eta_i$ and $K_m$ are arbitrary locally analytic functions of their arguments.
\looseness=-2

The substitution of (\ref{nshock}) into the equation
$u_t=F$ with $F$ of the form (\ref{nshockxt})
reduces this equation to the following system of ODEs:
$$
{\ds\frac{d c_i}{d t}}=c_i \eta_i
\biggl({\ds\frac{c_1^{k_{N+1}-k_{N}}}
{c_N^{k_{N+1}-k_{1}}}},\dots,
{\ds\frac{c_{N-1}^{k_{N+1}-k_{N}}}{c_N^{k_{N+1}-k_{N-1}}}}\biggr),
\quad i=1,\dots,N.
$$

If $\zeta_i=-h_i \sum_{j=0}^{m}\alpha_j (k_i^j-k_{N+1}^j)$, where
$\alpha_j$ are arbitrary constants, then (\ref{shockf}) represents
the most general $F$ of order $n$ such that the equation $u_t=F$
admits an $N$-shock solution of the form \cite{fliu}\looseness=-1
\be\label{shock3} u= -D\biggl(\ln\biggr(\sum\limits_{i=1}^{N+1}A_i
\exp\biggl(k_i x-t\sum\limits_{j=0}^{m}\alpha_j
k_i^j\biggr)\biggr)\biggr), \ee where $A_i$ are arbitrary constants.

In other words, the class of evolution equations
admitting the $N$-shock solutions of the form (\ref{shock3})
contains not only the equation \cite{fliu}
$$
u_t= \sum\limits_{j=0}^{m}\alpha_j D(D-u)^j(u)
$$
but infinitely many other equations $u_t=F$ as well.
In particular, if the order $n$ of $F$ is greater than $N-2$, then
Theorem~\ref{an} implies that the corresponding $F$'s have the form
$$
\ba{l}
F= \sum\limits_{i=1}^N \sum\limits_{r=0}^{m}\alpha_r
(k_i^r-k_{N+1}^r) {\ds\frac{\tilde
h_i\sum\limits_{j=1}^{N+1}(k_i-k_j)\tilde h_j}
{\biggl(\sum\limits_{q=1}^{N+1}\tilde h_q\biggr)^2}}\\
+\sum\limits_{m=0}^{n-N}
D^m(Q) K_{m}(x,t,u,u_1,\dots,u_{N-1}, Q, D(Q),\dots,
D^{n-N}(Q)). \ea
$$

\section{Conclusions and discussion}
In Theorem~\ref{an} of the present paper we completely characterized
locally analytic evolution systems~(\ref{eveq}) of a given order
$n\geq k-1$
that admit a generalized conditional
symmetry $\Q(x,t,\allowbreak\bu,\dots,\bu_k)\p/\p \bu$ with
a given analytic $\bi{Q}$ of the form (\ref{ndeg}).
In particular, for the case of linear GCS, i.e.,
when $\bi{Q}$ is linear in $\bi{u}_j$ for all $j$,
the right-hand sides $\F$
of such systems (\ref{eveq}) are given by (\ref{symlin}).

The results of the present paper are somewhat easier to use in
applications than the formulas found by us earlier in \cite{as-jpa}.
For instance, the usage of Theorem~\ref{an} enabled us to find the
explicit form of {\em all\/} locally analytic evolution equations of
order $n \geq N-1$ that admit $N$-shock (and, more broadly,
$N$-shock type) solutions for a given number $N$, and thus
generalize the results of Fokas and Liu \cite{fliu}.
\looseness=-1


Moreover, Theorem~\ref{an} can be employed for the classification of
exactly solvable initial value problems along the lines of
\cite{zbh, zhd3} and for finding evolution equations whose
symmetries are compatible with prescribed boundary conditions, and
one can find exact solutions for the corresponding boundary value
problems, cf.\ \cite{aggh} and discussion in \cite{as-jpa}. We plan
to address these topics elsewhere. \looseness=-1

\section*{Acknowledgments}
I acknowledge with gratitude the support from the Czech Grant Agency
(GA \v CR) under grant No.~201/04/0538 and the Ministry of
Education, Youth and Sports of Czech Republic (M\v SMT \v CR) under
grant~MSM:J10/98:192400002.


\end{document}